\documentclass[reprint]{revtex4-1}
\usepackage{hyperref}
\usepackage{graphicx}
\usepackage{dcolumn}
\usepackage{bm}
\usepackage[utf8]{inputenc}
\usepackage[T1]{fontenc}
\usepackage{mathptmx}
\usepackage{etoolbox}
\usepackage{multirow}
\usepackage{color}
\usepackage{xcolor}

\usepackage{graphicx}
\usepackage{amssymb}
\usepackage{textcomp}
\usepackage{float}
\usepackage{natbib}
\usepackage{ragged2e}
\usepackage{color,soul}
\usepackage{array}
\usepackage{dcolumn}
\usepackage{url}
\usepackage{multirow}
\usepackage{appendix}
\usepackage[normalem]{ulem}
\usepackage{soul}
\usepackage{caption}
\usepackage{subcaption}
\usepackage{gensymb}

\begin{document}

\preprint{AIP/123-QED}

\title[]{Formation of ultracold triatomic molecules by electric microwave association}

\author{Baraa Shammout}
\email{shammout@iqo.uni-hannover.de}
\affiliation{Institut für Quantenoptik, Leibniz Universität Hannover, 30167 Hannover, Germany}

\author{Leon Karpa}
\affiliation{Institut für Quantenoptik, Leibniz Universität Hannover, 30167 Hannover, Germany}

\author{Silke Ospelkaus}
\affiliation{Institut für Quantenoptik, Leibniz Universität Hannover, 30167 Hannover, Germany}

\author{Eberhard Tiemann}
\affiliation{Institut für Quantenoptik, Leibniz Universität Hannover, 30167 Hannover, Germany}

\author{Olivier Dulieu}
\email{olivier.dulieu@universite-paris-saclay.fr}
\affiliation{Universit$\acute{\text e}$ Paris-Saclay, CNRS, Laboratoire Aim$\acute{\text e}$ Cotton, Orsay, 91400, France}

\date{\today}

\begin{abstract}


A theoretical model is proposed for the formation of ultracold ground-state triatomic molecules in weakly bound energy levels. The process is driven by the electric component of a microwave field, which induces the association of an ultracold atom colliding with an ultracold diatomic molecule. This model is exemplified using $^{39}$K atoms and $^{23}$Na$^{39}$K molecules, both in their ground states, a scenario of experimental relevance. The model assumes that the dynamics of the association are dominated by the long-range van der Waals interaction between $^{39}$K and $^{23}$Na$^{39}$K. The electric microwave association mechanism relies on the intrinsic electric dipole moment of $^{23}$Na$^{39}$K, which drives transitions between its lowest rotational levels ( $j$=0 and $j$=1). The energies of the uppermost triatomic energy levels are computed by numerically solving coupled Schrödinger equations using the Mapped Fourier Grid Hamiltonian method. Measurable association rates are derived within the framework of a perturbative approach. This method of electric microwave association provides an alternative to atom-molecule association via magnetic Feshbach resonances for forming ultracold, deeply bound triatomic molecules, and is applicable to a wide range of polar diatomic molecules.
\end{abstract}

\maketitle 

\section{Introduction}
\label{sec:intro}

Ultracold quantum-degenerate gases have been a flourishing field of research in atomic and molecular physics for three decades. The ultralow kinetic energy $E_c$ of the particles in ultracold dilute gases (corresponding to temperatures $T=E_c/k_B << 1$~mK, with typical number densities of $10^9-10^{12}$~cm$^{-3}$), allows for an exquisite level of control of their dynamics using external electromagnetic fields. The founding example is the exposure of ultracold dilute atomic gases to a static magnetic field of moderate strength (typically from a few tens to a few hundreds gauss), in order to control the scattering length of a colliding atom pair via magnetic Feshbach resonances (MFR) \cite{tiesinga1993,kohler2002}, a key step for the experimental achievement of ultracold degenerate gases. The subsequent formation of so-called Feshbach molecules using a suitable ramped magnetic field represents the initial step for the formation of ultracold diatomic molecules in their absolute ground state, a routine experimental procedure nowadays \cite{ni2008}. A combination of a static magnetic field and a microwave field (MW) has been found to improve the fine-tuning of Feshbach molecules formation \cite{voges2019,voges2020a}. The association of a pair of atoms into a loosely bound molecule with a MW field has been observed \cite{maury2023}, which paved the way for the observation of mw-induced MFRs \cite{papoular2010}. 

More complex ultracold systems have recently benefited from such control capabilities. MFRs have been detected in collisions between ultracold atoms and molecules \cite{wang2021,son2022,park2023,meyer2025}, resulting in the formation of loosely bound ultracold triatomic molecules \cite{yang2022a,yang2022b}, and in the first signal of (optical) photoassociation (PA) resonances in ultracold trimers \cite{cao2024}. When ultracold molecular species possess a permanent electric dipole moment (PEDM) in their own frame, thus designated as polar molecules, their mutual long-range dipole-dipole interactions (LR-DDI) can be engineered using strong electric fields (larger than 10~kV/cm) to enhance their anisotropic pattern \cite{quemener2010b,matsuda2020}. LR-DDI can also be tuned with a MW field to acquire a repulsive character, thus preventing trapped molecules from colliding with each other in the so-called sticky regime, leading to their loss from the trap \cite{lassabliere2018,karman2018,anderegg2021,schindewolf2022,bigagli2023}. This enabled the first observation of Bose-Einstein condensation of ground-state molecules \cite{bigagli2024}. MW fields are a versatile approach in a similar context, allowing the observation of halo trimer levels of Na$_2$K through MW association \cite{chuang2024}, of field-linked bound levels of pairs of diatomic molecules \cite{avdeenkov2003,chen2023}, and of macro-dimer molecules created from pairs of Rydberg atoms \cite{bai2024}.

In this paper, we propose to extend the study of MW-induced association to a pair composed of an atom and a polar molecule. In contrast to \cite{papoular2010}, relying on the magnetic component of the MW-field to induce association, its electric component can induce transitions between rotational states by stimulating the PEDM of the polar molecule. These transitions are expected to be stronger than the magnetic transitions in atoms and should then open up a convenient way to populate loosely bound levels of the ultracold atom-diatom complex without relying on MFR. Combined with recent advances in knowledge on atom-molecule PA \cite{perez-rios2015,elkamshishy2022,shammout2023,cao2024}, this approach could represent a suitable initial step toward the formation of ultracold triatomic molecules in their absolute ground state. 

We first recall the main features of the long-range PA model of \cite{shammout2023}, which we adapt to the particular case of the interaction between ultracold ground-state $^{39}$K atoms and ultracold $^{23}$Na$^{39}$K molecules in the lowest vibrational level (v=0) of their electronic ground state $X^1\Sigma^+$, prepared at the rotational level $j=0$. In the following, we omit the atomic mass numbers in the notation. The Electric Microwave Association (EMWA) is defined as:
\newpage
\begin{eqnarray}
    \textrm{K}(4^2S_{1/2})+\textrm{Na}\textrm{K}(X^1\Sigma^+,v=0, j=0)+ h\nu \\ \nonumber
    \rightarrow \textrm{K} \cdots \textrm{NaK},
\label{eq:emwa}
\end{eqnarray}
where the MW photon has a frequency $\nu$ sufficiently detuned from the one of the pure rotational transition $j=0 \rightarrow j=1$ in NaK($X^1\Sigma^+,\textrm{v}=0$), while K$\cdots$NaK refers to the loosely bound triatomic complex in its $^2A'$ electronic ground state \cite{shammout2023}. In the following, we will invoke only the $j$ quantum number to characterize the NaK molecular level. We solved coupled Schr\"odinger equations to determine the energies and wave functions of loosely bound levels of the K$\cdots$NaK complex below the K + NaK($j=0$) limit, as well as the energies and widths of the predissociating resonances located between K + NaK($j=0$) and K + NaK($j=1$). We evaluated in the weak MW field limit their association rate,  which is found to be measurable under realistic experimental conditions.

\section{Electric Microwave Association mechanism}
\label{sec:emwa}

\begin{figure}
    \centering
\includegraphics[scale=1]{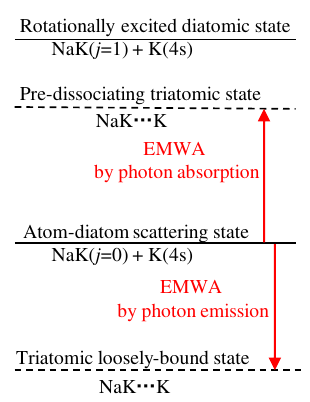}
    \caption{Scheme of the proposed Electric MicroWave Association (EMWA). The scattering complex K + NaK($j=0$) either absorbs a MW photon toward a K$\cdots$NaK predissociating resonance (upward arrow), or stimulates a photon down to a K$\cdots$NaK weakly bound level (downward arrow). }
    \label{fig:MW_PA}
\end{figure}

Figure \ref{fig:MW_PA} displays two possible pathways for EMWA. During the collision of a ground-state K atom and a ground-state NaK molecule prepared in the $j=0$ level in free space, the scattering complex can either absorb a MW photon toward a predissociating resonance of the K$\cdots$NaK complex located above the K + NaK($j=0$) limit, or emit a photon by stimulated emission down to a weakly bound level of K$\cdots$NaK below that limit. We note that the latter process has been envisioned for atom-atom PA \cite{juarros2006}, or for laser-assisted self-induced Feshbach resonances between atoms \cite{devolder2019}, revealing, however, a poor efficiency due to a weak transition dipole moment (TDM). In contrast, the proposed process benefits from the strong TDM of the rotational transition in the NaK molecule. In the following, we consider the energy reference at the K + NaK($j=0$) limit. Predissociating resonances will be labeled with a positive index $n$ and positive energies $E_n$, while weakly bound levels will be labeled with a negative index $-n$ with negative energies $E_{-n}$.

Full quantum scattering calculations for such a process are almost impossible, as they require complete knowledge of the potential energy surface of the NaK$_2$ molecule and the computation of the resulting dense energy level spectrum. Instead, we conduct a quantitative study analogous to our earlier investigation \cite{shammout2023}, involving only the long-range potential energy surface of the ground-state K$\cdots$NaK complex, and restricting the dynamics to the rotational degree of freedom of NaK(v=0) and the overall rotation of the K$\cdots$NaK pair characterized by the partial wave $\ell$.  \\

\textit{Long-range interactions between K and NaK.}
The long-range potential energy surface (PES) for the electronic ground state of K$\cdots$NaK is calculated in Jacobi coordinates ($R$, $r$, $\theta)$ with $r$ being the NaK bond length, $R$ the distance between the K atom and the center of mass of NaK, and $\theta$ the angle between the axis containing $r$ and the one containing $R$. We varied $R$ over the [30~a.u.-160~a.u.] range with unequal step sizes (29 points) and $r$ in the [5.804~a.u.-7.436~a.u.] range with a step size of 0.204~a.u. (9 points). The $r$ interval is chosen to cover the extension of the vibrational motion in NaK(v=0). The Jacobi angle $\theta$ is varied in the [0\degree-180\degree] range with an increase of 10\degree (19 points). In total, the mesh grid consists of $29\times 9 \times 19 = 4959$ points. The PES is calculated with the MOLPRO package using the MRCI method (see Appendix \ref{Appendix: Ground state long-range interaction}) \cite{werner2012,shammout2023}. Beyond $R=160$~a.u., the PES is extrapolated according to the long-range formula of Eq.\ref{eq:Long_range_expansion} in the Appendix.\\

\textit{Basis set for the coupled channels.}
We define the coupled asymptotic basis set $|j,\ell,J \rangle $ for a rotating NaK with quantum number $j$ associated with the rotational angular momentum $\hat{j}$, and a ground-state K atom (neglecting its internal angular momentum), with their relative rotation described by the quantum number $\ell$ associated to the angular momentum $\hat{\ell}$, \textit{i. e.} the partial wave. The total angular momentum $\hat{J}=\hat{j}+\hat{\ell}$ and the parity $P=(-)^{j+\ell}$ are conserved in the absence of external fields and if the hyperfine structure of the colliding partners is neglected.  We consider the entrance channel K + NaK($j=0$) in the $s$-wave ($\ell=0)$ with positive parity $P=1$. The final states (according to the electric dipole selection rule) of an association process with $J=1$ and negative parity $P=-1$ in the space-fixed frame are expanded in the basis vectors correlated with the dissociation limits K + NaK($j$) up to $j_{max}=6$ and $\ell_{max}=7$ (Table \ref{tab:Basis vectors-SF}). The size of this basis set yields satisfactory converged level energies over a range of about $h\times$6.7~GHz (namely, including the energy interval ($h\times$5.74~GHz) between the $j=0$ and $j=1$ NaK levels) below and above the K + NaK($j=0$) limit (see Appendix \ref{Appendix: convergence}). The coupled-channel calculations are performed using the Mapped Fourier Grid Hamiltonian (MFGH) method \cite{kokoouline1999}, which also allows the identification and characterization of predissociating resonances in the dissociation continuum between the K + NaK($j=0$) and K + NaK($j=1$) limits by the stabilization method (see Appendix \ref{Appendix: convergence}). \\
\begin{table}
\caption{\label{tab:Basis vectors-SF} Basis vectors $|j,\ell,J \rangle $ for $J=1, j_{max}=6,\ell_{max}=7$ and $P=-1$ in the space-fixed frame, correlated to the dissociation limits K + NaK($j$) with energies $\epsilon_j$ relative to K + NaK($j=0$).}
\begin{tabular}{c|c|c|c}
\hline 
\hline
$J$&$j$ & $\ell$ & $\epsilon_j/h$ (GHz)   \\ \hline
1&0 & 1 & 0\\
1&1 & 0 & 5.74 \\
1&1 & 2 & 5.74\\
1&2 & 1 & 17.23\\
1&2 & 3 & 17.23\\
1&3 & 2 & 34.46\\
1&3 & 4 & 34.46\\
1&4 & 3 & 57.44 \\
1&4 & 5 & 57.44\\
1&5 & 4 & 86.15 \\
1&5 & 6 & 86.15 \\
1&6 & 5 & 120.61 \\
1&6 & 7 & 120.61  \\
\hline 
\hline
\end{tabular}
\end{table}
     
\textit{Matrix elements of the potential operator.}
These $R$-dependent matrix elements are obtained after numerical integration on $r$ and $\theta$ of the PES depending on $R$, $r$ and $\theta$ \cite{shammout2023}. The diagonal matrix elements $V^{J=1}_{j \ell, j \ell}$ and the off-diagonal matrix elements $V^{J=1}_{j \ell, j' \ell'}$ of the long-range potential operator in the space-fixed frame are obtained by a frame transformation  \cite{launay1976,hutson1991,lara2015} of the diagonal matrix elements $V^{J=1}_{j k_j, j k_j}(R)$ and the off-diagonal matrix elements $V^{J=1}_{j k_j, j' k_{j'}}(R)$ in the body-fixed frame, with $k_j=k_{j'}$, and $k_j$ being the projection of $J$ on the axis containing $R$ ($k_j=-1,0,1$ for $J=1$). \\
     
\textit{Model for the potential operator at short-range.}
In the short range ($R<30$~a.u.) we match the $V^{J=1}_{j k_j, j k_j}(R)$ functions to the effective potentials $V^{\textrm{sr}}(R)$ of the Lennard-Jones form
\begin{equation}
      V^{\textrm{sr}}_{J j \ell}(R)=D^{\textrm{sr}}(C^{\textrm{sr}}_{J j \ell}/R^6)[(C^{\textrm{sr}}_{J j \ell}/R^6)-1]+E_{j \ell}(R), 
\label{eq:LJ_pot}
\end{equation}
where $E_{j\ell}(R)=\epsilon_j+ \ell(\ell+1)/(2\mu R^2)$, and $\epsilon_j$ is the rotational energy of the diatom (see Table \ref{tab:Basis vectors-SF}. We kept $D^{\textrm{sr}}=0.175$~a.u. constant and varied $C^{\textrm{sr}}_{J j \ell}$ to smoothly connect the effective diagonal long-range matrix elements $V^{J=1}_{j \ell, j \ell}(R) +E_{j \ell}(R)$ to the short-range potentials $V^{\textrm{sr}}_{J j \ell}(R)$. Thus we will not obtain predictions of the EMWA spectrum but typical levels and their wave function to estimate rate intensities and, therefore, the feasibility of such an experiment.\\

\textit{Weakly-bound levels of the K$\cdots$NaK complex.}
The energies of weakly bound levels ($E_{-n}$) and predissociating levels ($E_{n}$) of the K$\cdots$NaK complex for $J=1$ and $P=-1$, calculated with the MFGH method, are presented in Table \ref{tab:Eigenvalues}, over an energy range of about $h \times6.7$~GHz above and below the K + NaK($j=0$) limit. The corresponding wave functions $|n, J\rangle \equiv \Psi_f^J(R,E_n)$ are expressed as  
\begin{equation}
    |n, J\rangle= \sum_{j\ell} \psi^{J}_{j\ell,n}(R) |j,\ell,J \rangle ,
\end{equation}
so that the squared components integrated over $R$ between the boundaries $R_{\textrm{min}}$ and $R_{\textrm{max}}$ implemented in the MFGH method
\begin{equation}
\alpha^{J}_{j\ell,n}=\int_{R_{\textrm{min}}}^{R_{\textrm{max}}}  \left[\psi^J_{j\ell,n}(R)\right]^2dR
\end{equation}
represent the partial norm of the eigenfunctions $|n, J\rangle$ on each basis vector $|j,\ell,J \rangle $, with $\sum_{j\ell} \alpha^{J}_{j\ell,n}=1$. Additional information on the convergence of these calculations with the size of the basis set is provided in the Appendix. To simplify the notation, we omit the index $(J=1)$ in the following.

We report in Table \ref{tab:Eigenvalues} the partial norms for the three lowest asymptotic basis vectors, the sum of which represents more than 60\% of the total norm in this energy range. We found five weakly bound levels, representing a density of states consistent with the one obtained from the semiclassical model described in \cite{christianen2019b} (See Eq.30 therein). Starting from K + NaK($j=0$) and assuming  $s$-wave collision ($\ell=0$), the levels with a significant component $(\alpha^J_{j=1,\ell=0})$, namely $n=-5,-2$, are efficiently excited by the MW radiation relying on the PDM of NaK. The quantum number $\ell$ is unchanged for such electric dipole-allowed transitions. 

We determine the energy of three predissociating resonances $n=1,2,3$ between the K + NaK($j=0$) and K + NaK($j=1$) limits. These levels can be excited with the MW field due to their significant partial norm $(\alpha^J_{10, n})$. The height of the rotational barrier ($h\times 3.57$~MHz in agreement with the value $h\times 3.69$~MHz reported in \cite{zuchowski2013}) associated with the $p$ wave ($\ell=1$) in the entrance channel K + NaK($j=0$) is much lower than $E_1,E_2,E_3$, so these resonances do not correspond to shape resonances. Instead, they dissociate toward the continuum of K+NaK($j=0$), via the $\ell=1$ channel due to rotational coupling. Thus, only levels with a low partial norm $(\alpha^J_{01, n})$ are expected to have a sufficiently long lifetime (and thus a low predissociation width $\Gamma_n$) to be detected, namely $n=1$ here. The upper limits on $\Gamma_n$ displayed in Table \ref{tab:Eigenvalues} are the result of convergence tests based on the stabilization method described in the appendix. These widths could be estimated with better accuracy by lowering the energy step size in the stabilization method.
\\

\begin{table}
\caption{\label{tab:Eigenvalues} Computed eigenvalues $E_n$ for the  K$\cdots$NaK complex with $J$=1, $P=-1$, below and above the dissociation limit K + NaK($j=0$). The partial norms $\alpha^J_{j\ell},n$ are reported for the three lowest basis vectors in Table \ref{tab:Basis vectors-SF}. An upper bound for the predissociation width $\Gamma_n$ is given for $n=1,2,3$.}
\begin{ruledtabular}
\begin{tabular}{l|c|c|c|c|c}
$n$ & $E_{n}/h$ (GHz) & $\alpha^J_{01, n}$ & \multicolumn{1}{c|}{$\alpha^J_{10, n}$} & $\alpha^J_{12, n}$ & $\Gamma_n$ (MHz) \\ \cline{1-6}
-5  & -6.627& 0.04& \textbf{0.17}& 0.63 & -    \\
-4  & -6.352& 0.66& \textbf{0.03}& 0.07 & -   \\
-3  & -2.390& 0.79& \textbf{0.01}& 0.03 & -    \\
-2  & -1.088& 0.09& \textbf{0.69}& 0.15 & -    \\
-1  & -0.292& 0.88& \textbf{0.05}& 0.01 & -    \\
1   & 3.605 & 0.03& \textbf{0.57}& 0.03 & <107  \\
2   & 5.119 & 0.52& \textbf{0.24}& 0.01 & <370  \\
3   & 5.699 & 0.36& \textbf{0.55}& 0.02 & <140
\end{tabular}
\end{ruledtabular}
\end{table}

\begin{figure}
    \centering
\includegraphics[scale=0.65]{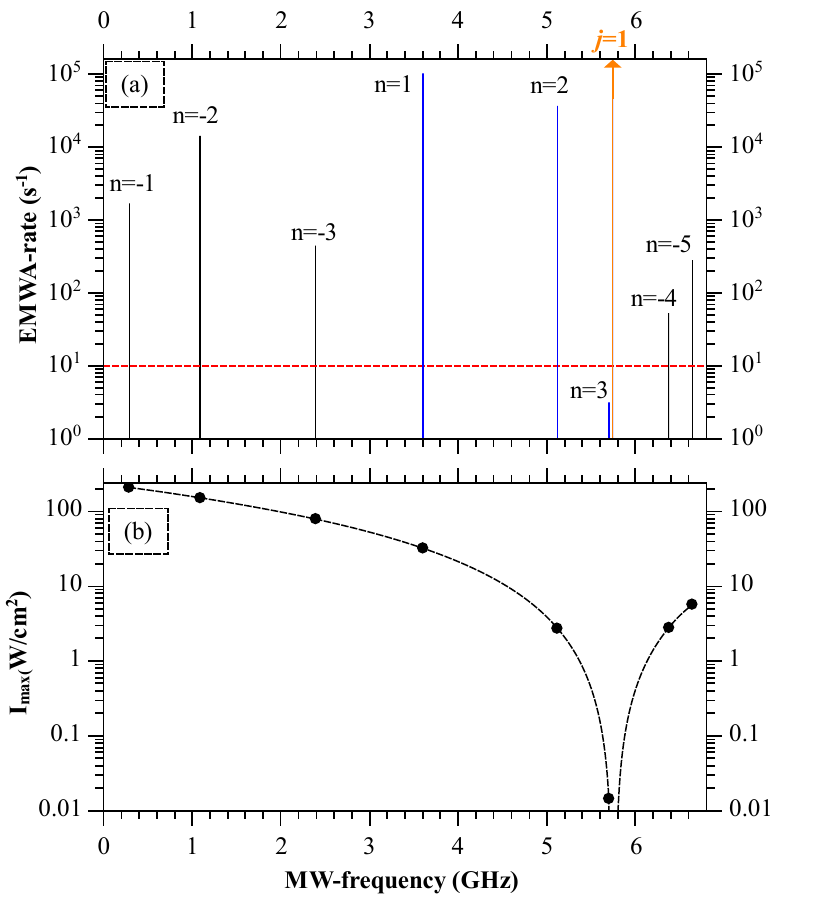}
    \caption{(a) Calculated EMWA rate per diatomic molecule of K and NaK($j=0$) at a collisional energy of $E_r=k_B \times 200$~nK and molecular density $n_{\textrm{NaK}}=10^{12}$~cm$^{-3}$, as a function of the MW frequency. The vertical black lines represent loosely bound levels below the threshold K + NaK($j=0$), while blue lines correspond to predissociating resonances. All resonances are labeled according to Table \ref{tab:Eigenvalues}. The orange line marks the energy of the $j=1$ rotational state of NaK. The red dashed line represents typical trap loss rate. (b) the corresponding microwave intensity threshold for a weak perturbation, considering $2\pi\times\Delta=10~\Omega$. Note that the rates reported in panel (a) are computed at each of these different intensities, so that the height of the peaks cannot be compared with each other to look for feasible detection.  }
    \label{fig:PA-rate}
\end{figure}

\textit{Weak-field limit for EMWA.}
We take into consideration the possible Rabi oscillations due to MW-induced transitions in the diatom NaK between its rotational energy levels $j=0$ and $j=1$ (separated by $h \nu_{j=0 \to j=1}=h \times 5.74$~GHz from the present calculation, close to its experimental value $h \times5.6963$~GHz \cite{yamada1992}). We desire a weak perturbation of the NaK molecule by the MW field. Thus, the MW intensity is limited to ensure that the diatomic Rabi frequency $\Omega$ is much smaller than the MW detuning $\Delta=|\nu_{j=0 \to j=1} - f_{\textrm{mw}}|$. The expression for the MW intensity as a function of $\Omega$ in SI units is written as $I$(W/m$^2$)$=\epsilon_0 c (\hslash \Omega/d_{\textrm{NaK}})^2/2$, where $\epsilon_0$ is the vacuum permittivity, $c$ is the speed of light, and $d_{\textrm{NaK}}$ is the permanent dipole moment of NaK. For the condition $\Omega=2\pi \Delta/10$, we calculate $I_{\textrm{max}}$ as a function of the MW frequency $f_{MW}= \nu_{j=0 \to j=1} \pm \Delta$ as shown in Fig.\ref{fig:PA-rate}(b).\\

\textit{EMWA rate in the perturbative regime.}
In a perturbative regime, the rate of EMWA $R_{if}$   \cite{shammout2023,pillet1997,cote1998,perez-rios2015,elkamshishy2022} is inversely proportional to the temperature $T$ and directly proportional to the squared PEDM of the diatom ($d_{\textrm{NaK}}=2.72(6)$~D \cite{gerdes2011}), to the density of diatomic molecules (considering the diatoms being the minority ensemble inside a cloud of atoms), to the MW field intensity $I_{\textrm{MW}}$ (as long as $I\le I_{\textrm{max}}$), and to the squared overlap
\begin{equation}
|S_{if}(E_r)|^2 = \left[\int_{R_{\textrm{sr}}}^{R_{\textrm{max}}}  \psi^J_{j\ell,n}(R)\xi_i(R, E_r)dR\right]^2
\end{equation}
 between the radial components $\psi^J_{j\ell,n}(R)$ of the total wave function $\Psi_f^J(R,E_n)$ (due to the selection rules we consider only the component with $j=1,\ell=0$) and the continuum radial wave function $\xi_i(R, E_r)$ of the the K$\cdots$NaK complex at energy $E_r$, which is obtained as in our former paper \cite{shammout2023}. Note that the calculation of spatial overlap is restricted to the long-range region between $R_{\textrm{sr}}$=30~a.u. and $R_{\textrm{max}}$=1000~a.u. relevant for the present study. \\



\textit{Experimental feasibility.}
For typical experimental conditions of molecular density $n_{\textrm{NaK}}=10^{12}$ cm$^{-3}$ and temperature $T=E_r/k_B=200$~nK, we present in Fig.\ref{fig:PA-rate} (upper-panel) the calculated EMWA rates of K$\cdots$NaK in s$^{-1}$ per diatomic molecule NaK as a function of the MW frequency $f_{\textrm{MW}}$ in GHz up to 6.63~GHz for maximum microwave intensities $I_{\textrm{max}}$($f_{\textrm{MW}}$) up to 209~W/cm$^2$ according to Fig.\ref{fig:PA-rate}(b) for the weak-field approximation. For a feasible experimental observation, the EMWA rate must be significantly higher than the typical loss rate from the molecular trap without the MW-field (around 10~s$^{-1}$ \cite{voges2020}, see the horizontal dashed line in Fig.\ref{fig:PA-rate}(a)). Bound levels $n=-1, -2, -3,-4, -5$ have rates which are high enough such that they should already be detectable with intensities around 10~W/cm$^2$. The excitation of the predissociating resonances is also feasible depending, however, on their width. The present model yields the $n=1,3$ resonances with a relatively narrow width, while the $n=2$ width could be large to consider for detection, 

\section{Conclusion}

We have introduced a theoretical proposal for the formation of ultracold ground-state loosely bound triatomic molecules via EMWA of a ground-state atom with a ground-state polar diatomic molecule prepared in an ultracold gas mixture. We have studied as an example of experimental relevance the association of NaK($X^1\Sigma^+$, $v=0, j=0$) with K($4s$). In the regime of a weak MW field. Our calculations have shown that several loosely bound triatomic states can be formed under feasible experimental requirements. Higher MW power would disrupt the present perturbative description, but the association itself could still work, potentially showing broadened or shifted resonances. Our proposal relies on the intrinsic PEDM of diatomic molecules and is therefore broadly applicable. It is also particularly well-suited for molecules with small magnetic moments, such as $^1\Sigma^+$ molecules, for which Feshbach resonances are typically weak \cite{wang2021,meyer2025}.

Future investigations will concern the following aspects and extensions of the present approach: (i) A more detailed description of the short-range interaction potential for the K$\cdots$NaK complex, including the first excited electronic state, to account for possible conical intersections and non-adiabatic effects; (ii) A full quantum mechanical treatment of the EMWA process, including the time evolution of the system under the influence of the microwave field, (iii) The inclusion of the fine and hyperfine structure of the atom and molecule in the calculations, (iv) The treatment of non-perturbative effects. These developments should help in the optimization of the MW field parameters (frequency, intensity) to maximize the EMWA signal and minimize losses induced by predissociation and off-resonant transitions. This work opens a new prospect for the creation and manipulation of ultracold triatomic molecules using microwave radiation, paving the way for more complex investigations in the field of ultracold quantum chemistry, for example, to the preparation of ground-state triatomic molecules.\\

\section{Acknowledgments }
B. S., L. K, and S.O. gratefully acknowledge financial support from 
Germany’s Excellence Strategy EXC-2123 QuantumFrontiers 390837967, the Deutsche 
Forschungsgemeinschaft (DFG)
through CRC 1227 (DQ-mat), project A03, and the European Research 
Council through the ERC Consolidator grant 101045075 TRITRAMO.

\newpage
 \clearpage
 \section{Appendix }
 \setcounter{figure}{0}
\counterwithin{figure}{subsection}

\subsection{Electronic structure calculation}
\label{Appendix: Ground state long-range interaction}

We calculated the long-range potential energy surface (PES) for the electronic ground state of the K$\cdots$NaK complex with the multiconfiguration reference internally contracted configuration interaction (MRCI) method \cite{werner1988} with Pople correction using effective core potentials (ECPs), core polarization potentials (CPPs) and the basis sets defined in our previous work \cite{shammout2023}. The initial guess for orbitals is obtained by the multiconfiguration self-consistent field (MCSCF) method \cite{werner1985}. 
 
\begin{figure}
    \centering
\includegraphics[scale=0.42]{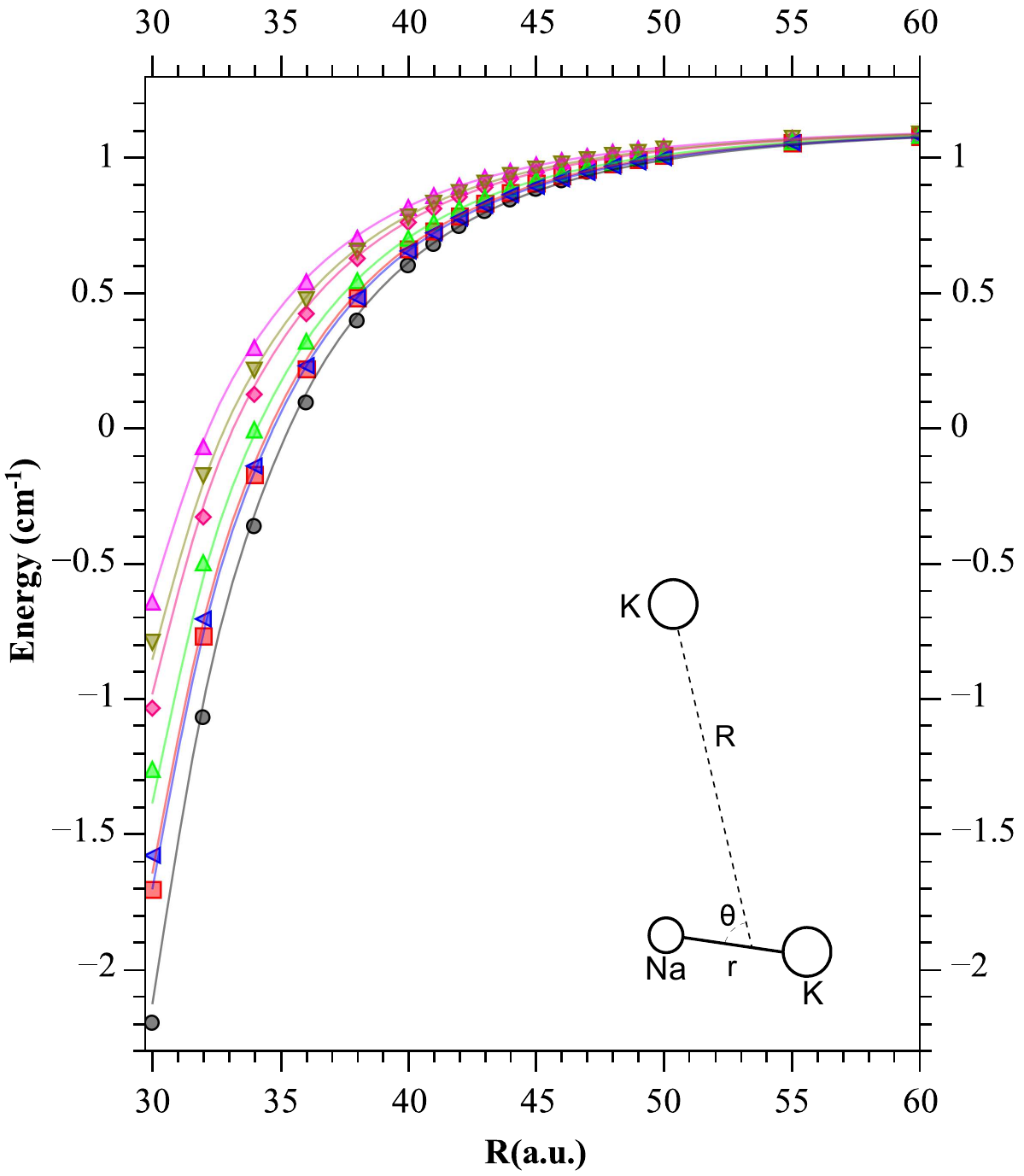}
    \caption{ \textit{Ab initio}  long-range potential energy curves (LR-PECs) of the electronic ground state of the K$\cdots$NaK complex in Jacobi coordinate $R$ at $r=6.62$~a.u. for different Jacobi angles $\theta$=0$\degree$ (black circles), 30$\degree$ (red rectangles), 60$\degree$ (pink diamonds), 90$\degree$ (dark purple triangles), 120$\degree$ (dark yellow down triangles), 150$\degree$ (green up triangles), 180$\degree$ (blue left triangles). Colored solid lines are the results of fit to Eq.\ref{eq:Long_range_expansion}. The energy reference is set to the dissociation limit NaK(X$^1\Sigma^+$ at $r_0$) + K($4s$), where $r_0=6.58$~a.u. is the equilibrium bond length of NaK from the present calculations (the experimental value is 6.612217(3)~a.u. \cite{yamada1992}).   }
    \label{fig:LR_PES_1A'}
\end{figure}

Figure \ref{fig:LR_PES_1A'} shows one-dimensional cuts through the long-range PES at $r=6.62$~a.u. for various Jacobi angles. To evaluate the long-range PES at any arbitrary point (with $R \ge 30$~a.u., $5.804 \textrm{ a.u.} \le r \le 7.436$~a.u. and 0\degree$ \le \theta \le 180$\degree), we fit the \textit{ab initio} points to the long-range multipolar expansion  
\begin{equation}
    V(R,r,\theta )=-\frac{C_6(r,\theta )}{R^6}-\frac{C_8(r,\theta )}{R^8}+E_{\infty}(r).
\label{eq:Long_range_expansion}
\end{equation}

\begin{figure}
    \centering
\includegraphics[scale=0.70]{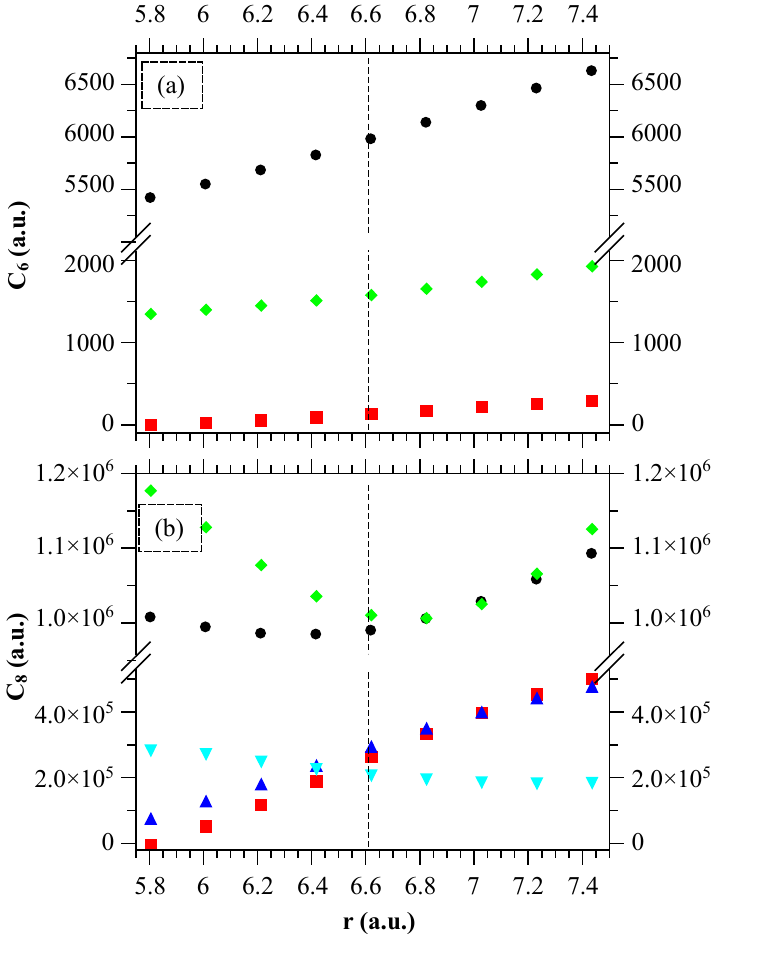}
    \caption{The isotropic and anisotropic $C_6^{(i)}$ (panel (a)) and $C_8^{(i)}$ (panel (b)) expansion coefficients for the triatomic complex NaK$\cdots$K in the electronic ground state as functions of the diatomic bond length $r$. Black circles: the isotropic $C_6^0$ and $C_8^0$. Red rectangles: the anisotropic coefficients $C_6^1$ and $C_8^1$. Green diamonds: $C_6^2$ and $C_8^2$. Blue triangles: $C_8^3$. Cyan down triangles: $C_8^4$. The vertical line indicates the experimental equilibrium bond length for NaK.  }
    \label{fig:C6}
\end{figure}

We obtain $C_6$ and $C_8$ long-range coefficients as functions of $r$ and $\theta$, where $E_{\infty}$($r$) is the asymptotic energy at a fixed $r$, representing the ground-state potential energy curve of the diatom. The channels with different $j$ and/or $\ell$ are coupled because of the anisotropy of the atom-diatom PES, which can be characterized by the expansion of the long-range coefficients in terms of Legendre polynomials $P_i(cos \theta)$ \cite{cvitas2006} according to the expressions
\begin{eqnarray}
C_6(r,\theta)=\sum_{i=0}^{2} C_6^{(i)} P_i(cos \theta), \\
C_8(r,\theta)=\sum_{i=0}^{4} C_8^{(i)} P_i(cos \theta).
\label{eq:longrangepot_iso_aniso}
\end{eqnarray}
Both expansions contain odd and even terms to account for the asymmetry of the K-NaK interaction with respect to $\theta=90$\degree. Figure \ref{fig:C6} presents the variations of these expansion coefficients with $r$. Our obtained value of the isotropic coefficient $C_6^{(0)}=5948$~a.u. at the NaK equilibrium distance is in good agreement with the one calculated in \cite{zuchowski2013} (5698~a.u.).  In both cases the influence of the asymmetry with respect to $\theta=90\deg$ is weak: the odd terms are small, while they are mandatory to properly fit the asymmetric variation of $C_6$ and $C_8$ . We see that the magnitude of the even terms of Eq. \ref{eq:longrangepot_iso_aniso} is noticeable, illustrating that the rotation of the diatom is indeed significantly hindered by the presence of the atom.

\subsection{Convergence of coupled channel calculations}
\label{Appendix: convergence}

To ensure the precision of our calculations, we first determine the length of the grid in the MFGH approach. The number of grid points in $R$ in the final calculations is 623, from $R_{\textrm{min}}=5.6$~a.u. up to  $R_{\textrm{max}}=1000$~a.u.. The eigenvalues of the Hamiltonian corresponding to bound levels are numerically converged to better than $10^{-5}$. Meanwhile, the energy position and widths of the predissociating resonances are obtained with an accuracy of around $h \times 100$~MHz, representing the energy step of the discretized dissociation continuum above NaK($j=0$) + K: this is the so-called stabilization method. A better accuracy in the numerical values could be obtained by increasing the value of $R_{\textrm{max}}$ resulting in a decreased energy step of the discretized continuum..

\begin{figure}
    \centering
\includegraphics[scale=0.70]{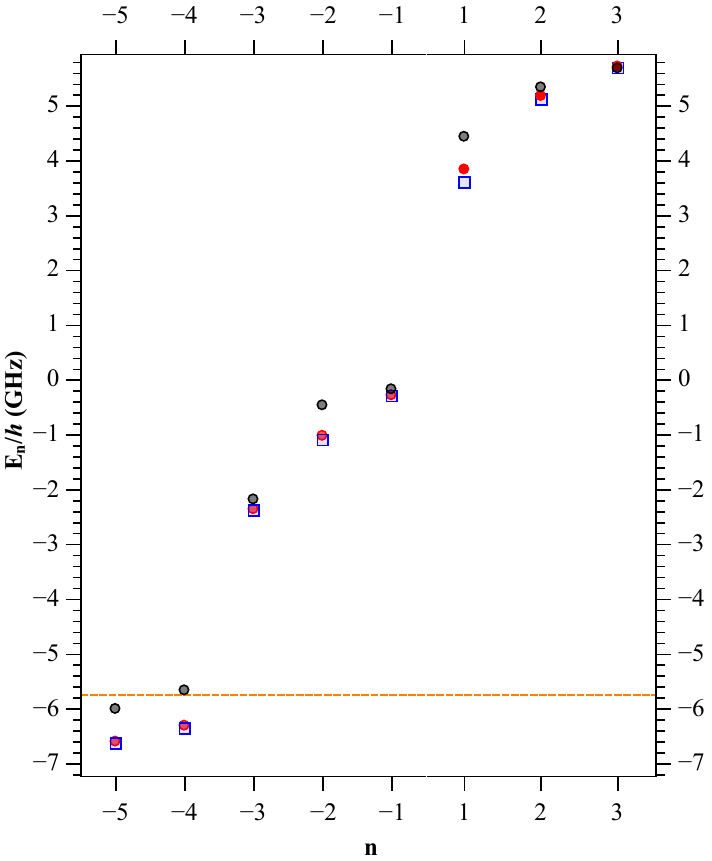}
    \caption{Computed eigenvalues $E_n$ for $J=1$ and $P=-1$ for different sizes of the basis set. Black circles: 5 channels. Red circles: 9 channels. Blue rectangles: 13 channels. The energy reference is set to the dissociation limit NaK($j=0$) + K($4s)$. The orange dashed line corresponds to $-\nu_{j=0 \to j=1}$. }
    \label{fig:convergence_energy}
\end{figure}

\begin{figure}
    \centering
\includegraphics[scale=0.70]{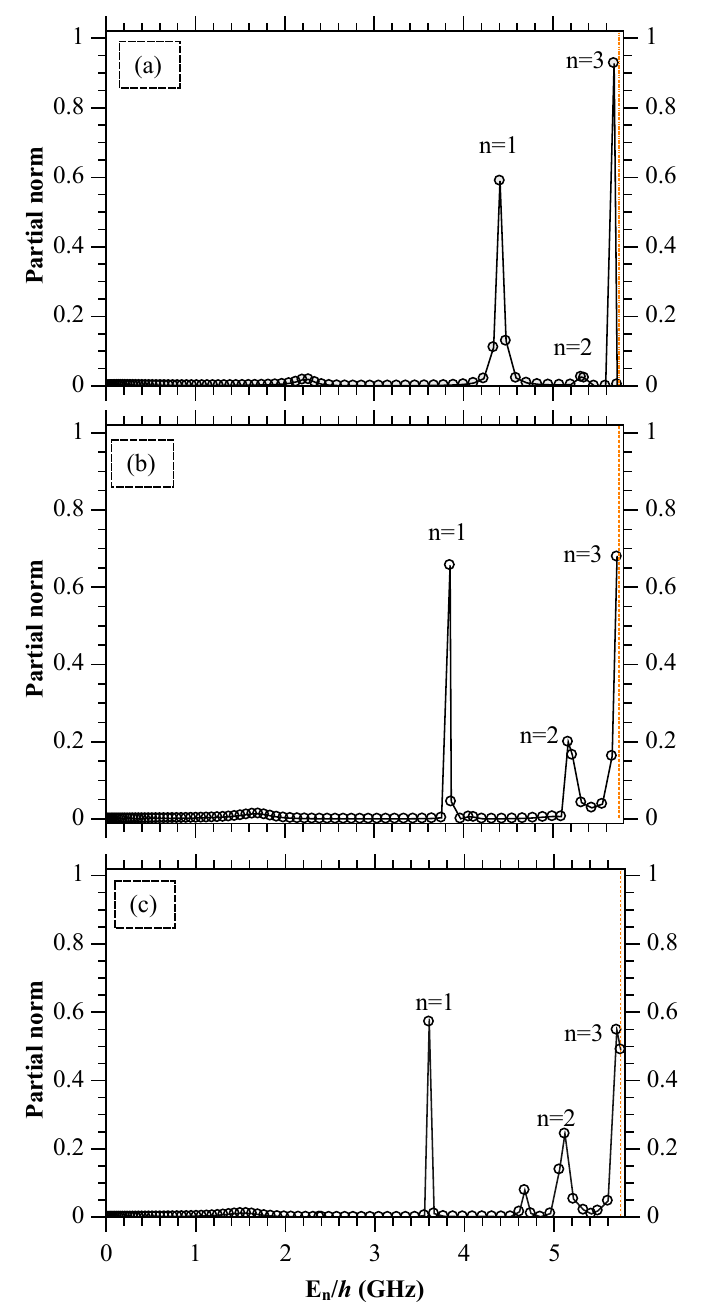}
    \caption{Eigenvalues yielded by the MFGH method between the $j=0$ energy (the origin of energies) and $j=1$ energy (marked with a vertical dotted line close to the right axis) of NaK, namely in the continuum of the K $\cdots$NaK($j=0$) complex for a basis set of 5 vectors (a), 9 vectors (b), and 13 vectors (c). The resonances labeled with $n$ are identified by the large partial norm $\alpha^J_{10, n}$ of the corresponding eigenvectors. All the other eigenvalues of the MFGH represent the discretized dissociation continuum.}
    \label{fig:convergence_prediss}
\end{figure}

We conducted convergence tests for eigenvalues with ($J=1$, $P=-1$) by systematically increasing the number of coupled channels (Fig.\ref{fig:convergence_energy}). We included up to 13 channels with $j_{max}=6, \ell_{max}=7$. For bound states ($n=-1, -2, -3, -4, -5$) below the NaK($j=0$) + K threshold, we achieved good convergence in both energy positions and partial norms. The convergence of the energies of predissociation resonances is not fully obtained for the three selected basis sizes: the shift of levels is no more than 250~MHz, for a basis size enlarged from 9 to 13 basis vectors. Figure \ref{fig:convergence_prediss} illustrates from panels (a) to (c) the convergence trend for the widths of the predissociating resonances with the size of the basis set, e.g. $n=1$ sharpens significantly by increasing the basis size. As invoked above with the stabilization method, the step size of the energy grid could be decreased to better describe the resonance profile, but it is sufficient for our current purpose of demonstrating the possibility to detect them.

\newpage
 \clearpage
 
	\bibliographystyle{unsrt}
	\bibliography{bibliocold,bibnote}

 \newpage
 \clearpage

\end{document}